# A dynamical measure of the black hole mass in a quasar 11 billion years ago

R. Abuter[1], F. Allouche[2], A. Amorim[3,4], C. Bailet[2], A. Berdeu[5], J.-P. Berger[6], P. Berio[2], A. Bigioli[7], O. Boebion[2], M.-L. Bolzer[8,9,19], H. Bonnet[1], G. Bourdarot[8], P. Bourget[20], W. Brandner[10], Y. Cao[8], R. Conzelmann[1], M. Comin[1], Y. Clénet[5], B. Courtney-Barrer[20,11], R. Davies[8], D. Defrère[7], A. Delboulbé[6], F. Delplancke-Ströbele[1], R. Dembet[5], J. Dexter[21], P.T. de Zeeuw[12], A. Drescher[8], A. Eckart[13,14], C. Édouard[5], F. Eisenhauer[8], M. Fabricius[8], H. Feuchtgruber[8], G. Finger[8], N.M. Förster Schreiber[8], P. Garcia[4,15], R. Garcia Lopez[24], F. Gao[13], E. Gendron[5], R. Genzel[8,16], J.P. Gil[20], S. Gillessen[8], T. Gomes[4,15], F. Gonté[1], C. Gouvret[2], P. Guajardo[20], S. Guieu[6], W. Hackenberg[1], N. Haddad[20], M. Hartl[8], X. Haubois[20], F. Haußmann[8], G. Heißel[5,26], Th. Henning[10], S. Hippler[10], S.F. Hönig[17], M. Horrobin[14], N. Hubin[1], E. Jacqmart[2], L. Jocou[6], A. Kaufer[20], P. Kervella[5], J. Kolb[1], H. Korhonen[10,20], S. Lacour[1,5], S. Lagarde[2], O. Lai[2], V. Lapeyrère[5], R. Laugier[7], J.-B. Le Bouquin[6], J. Leftley[2], P. Léna[5], S. Lewis[1], D. Liu[8], B. Lopez[2], D. Lutz[8], Y. Magnard[6], F. Mang[8,9], A. Marcotto[2], D. Maurel[6], A. Mérand[1], F. Millour[2], N. More[8], H. Netzer[22], H. Nowacki[6], M. Nowak[18], S. Oberti[1], T. Ott[8], L. Pallanca[20], T. Paumard[5], K. Perraut[6], G. Perrin[5], R. Petrov[2], O. Pfuhl[1], N. Pourré[6], S. Rabien[8], C. Rau[8], M. Riquelme[1], S. Robbe-Dubois[2], S. Rochat[6], M. Salman[7], J. Sanchez-Bermudez[25,10], D.J.D. Santos[8], S. Scheithauer[10], M. Schöller[1], J. Schubert[8], N. Schuhler[20], J. Shangguan[8], P. Shchekaturov[1], T.T. Shimizu[8], A. Sevin[5], F. Soulez[19], A. Spang[2], E. Stadler[6], A. Sternberg[22,23], C. Straubmeier[14], E. Sturm[8], C. Sykes[17], L.J. Tacconi[8], K.R.W. Tristram[20], F. Vincent[5], S. von Fellenberg[13], S. Uysal[8], F. Widmann[8], E. Wieprecht[8], E. Wiezorrek[8], J. Woillez[1], G. Zins[1]

[1] European Southern Observatory, Karl-Schwarzschild-Straße 2, 85748 Garching, Germany
[2] Université Côte d'Azur, Observatoire de la Côte d'Azur, CNRS, Laboratoire Lagrange, Nice, France
[3] Universidade de Lisboa - Faculdade de Ciências, Campo Grande, 1749-016 Lisboa, Portugal
[4] CENTRA - Centro de Astrofísica e Gravitação, IST, Universidade de Lisboa, 1049-001 Lisboa, Portugal
[5] LESIA, Observatoire de Paris, Université PSL, Sorbonne Université, Université Paris Cité, CNRS, 5 place Jules Janssen, 92195 Meudon, France
[6] Univ. Grenoble Alpes, CNRS, IPAG, 38000 Grenoble, France
[7] Institute of Astronomy, KU Leuven, Celestijnenlaan 200D, B-3001, Leuven, Belgium
[8] Max Planck Institute for extraterrestrial Physics, Giessenbachstraße 1, 85748 Garching, Germany
[9] Department of Physics, Technical University Munich, James- Franck-Straße 1, 85748 Garching, Germany
[10] Max Planck Institute for Astronomy, Königstuhl 17, 69117 Heidelberg, Germany
[11] Research School of Astronomy and Astrophysics, College of Science, Australian National University, Canberra, Australian Capital Territory, Australia
[12] Leiden University, 2311EZ Leiden, The Netherlands
[13] Max Planck Institute for Radio Astronomy, Auf dem Hügel 69, 53121 Bonn, Germany
[14] 1st Institute of Physics, University of Cologne, Zülpicher Straße 77, 50937 Cologne, Germany
[15] Faculdade de Engenharia, Universidade do Porto, rua Dr. Roberto Frias, 4200-465 Porto, Portugal
[16] Departments of Physics and Astronomy, Le Conte Hall, University of California, Berkeley, CA 94720, USA
[17] School of Physics & Astronomy, University of Southampton, Southampton, SO17 1BJ, UK
[18] Institute of Astronomy, Madingley Road, Cambridge CB3 0HA, UK
[19] Univ. Lyon, Univ. Lyon 1, ENS de Lyon, CNRS, Centre de Recherche Astrophysique de Lyon UMR5574, F-69230, Saint-Genis-Laval, France
[20] European Southern Observatory, Casilla 19001, Santiago 19, Chile
[21] Department of Astrophysical & Planetary Sciences, JILA, University of Colorado, Duane Physics Bldg., 2000 Colorado Ave, Boulder, CO 80309, USA
[22] School of Physics and Astronomy, Tel Aviv University, Tel Aviv 69978, Israel
[23] Center for Computational Astrophysics, Flatiron Institute, 162 5th Ave., New York, NY 10010, USA
[24] School of Physics, University College Dublin, Belfield, Dublin 4, Ireland

25 Instituto de Astronomía, Universidad Nacional Autóma de México, Apdo. Postal 70264, Ciudad de Méxixo 04510, Mexico
26 Advanced Concepts Team, European Space Agency, TEC-SF, ESTEC, Keplerlaan 1, 2201AZ, Noordwijk, The Netherlands

**Tight relationships exist in the local universe between the central stellar properties of galaxies and the mass of their supermassive black hole[1,2,3]. These suggest galaxies and black holes co-evolve, with the main regulation mechanism being energetic feedback from accretion onto the black hole during its quasar phase[4,5,6]. A crucial question is how the relationship between black holes and galaxies evolves with time; a key epoch to probe this relationship is at the peaks of star formation and black hole growth 8-12 billion years ago (redshifts 1-3)[7]. Here we report a dynamical measurement of the mass of the black hole in a luminous quasar at a redshift of 2, with a look back time of 11 billion years, by spatially resolving the broad line region. We detect a 40 micro-arcsecond (0.31 pc) spatial offset between the red and blue photocenters of the Hα line that traces the velocity gradient of a rotating broad line region. The flux and differential phase spectra are well reproduced by a thick, moderately inclined disk of gas clouds within the sphere of influence of a central black hole with a mass of $3.2 \times 10^8$ solar masses. Molecular gas data reveal a dynamical mass for the host galaxy of $6 \times 10^{11}$ solar masses, which indicates an under-massive black hole accreting at a super-Eddington rate. This suggests a host galaxy that grew faster than the supermassive black hole, indicating a delay between galaxy and black hole formation for some systems.**

SDSS J092034.17+065718.0 (hereafter J0920) is one of the most luminous quasars at *z ~ 2*, making it an attractive target for studies of supermassive black hole (SMBH) growth and its connection to host galaxy growth. Assuming the local broad line region (BLR) radius luminosity relationship[8] can be applied at high redshift, J0920 is then expected to have a large BLR. Given also its close proximity to a bright star and its bright Hα emission line redshifted into the *K*-band, we observed J0920 with GRAVITY+[9] at the Very Large Telescope Interferometer (VLTI), an upgrade to GRAVITY[10], using the new wide-field, off-axis fringe tracking mode (GRAVITY-Wide)[11].

From the raw GRAVITY+ frames, we extracted average differential phase curves of J0920 for each of the six baselines. For targets much smaller than the resolution limit, the differential phase is proportional to the displacement of the source photocenter along the baseline. We detect an "S-shape" differential phase signal in the longest baselines (Figure 1b and Extended Data Figure 1) characterizing a velocity gradient through the Hα line (Figure 1a) and suggesting a BLR dominated by rotation as found in local active galactic nuclei (AGN)[12,13,14].

We measure model-independent photocenters for the central 10 wavelength channels using all six baselines (Figure 1c) and observe a global East-West shift from the blue to the red wing of the line indicative of a velocity gradient. By binning all redshifted and blueshifted channels together, we measure an average separation between the two sides of $D_{photo} = 37\pm12$ µas (0.31±0.10 pc at *z* = 2.325) indicating a detection significance of 3-6σ (see Methods). Photocenter separations, however, can only provide at best a lower limit on the true BLR size given the unknown geometry in particular the inclination and opening angle. For these as well as determining the central SMBH mass, detailed kinematic modelling is needed.

We therefore simultaneously fit the six differential phase spectra and total flux spectrum with a kinematic model. The kinematic model consists of a distribution of independent clouds moving within the gravitational potential of the SMBH (Methods). The spectra are well fit by this model (reduced $\chi^2$ = 0.6) and the best fit is shown as the red curve in Figures 1a and 1b. Extended Data Table 1 reports the best fit parameters and their 68$^{th}$ percentile confidence intervals along with a brief description and the prior used.

We infer a mean Hα emitting BLR radius of $R_{BLR} = 40^{+20}_{-13}$ μas ($0.34^{+0.17}_{-0.11}$ pc) within a moderately inclined disk ($i = 32°^{+8}_{-7}$) that is oriented on-sky with a position angle, PA = $87°^{+19}_{-25}$. We further infer the BLR half opening angle to be $\theta_o = 51°^{+11}_{-13}$ which combined with the inclination is consistent with an unobscured quasar. We show an on-sky representation of the best fit BLR cloud distribution in Figure 1d.

Our measured radius is a factor of 2.25 smaller than what would be inferred from the local Hβ based radius luminosity relation[11] (see Figure 2 and Methods). Previous studies have actually measured up to a factor of 1.5 *larger* sizes for the Hα emitting region compared to Hβ[15,16,17] as expected for a radially stratified BLR and including optical depth effects[18]. This would only increase the tension between our spectro-interferometric size and luminosity-based size– although one should bear in mind that the latter is a "single epoch" method that uses only the line width of the BLR and the AGN luminosity, and so carries with it a large uncertainty

Our smaller size though is consistent with the results at lower redshift for the high luminosity quasars 3C 273 and PDS 456 observed with GRAVITY as well as reverberation mapping of high-Eddington ratio AGN[19,20,21,22]. Indeed, combining the bolometric luminosity of J0920 (log $L_{Bol}$ = 47.2 – 47.9 erg s$^{-1}$; see Methods) with our GRAVITY+ measured SMBH mass, we find an Eddington ratio, $L_{Bol}/L_{Edd}$ = 7 - 20 which supports previous observations that super-Eddington accreting quasars have smaller BLRs relative to the radius-luminosity relation. More generally, this is further an independent confirmation that super-Eddington quasars exist using a highly accurate SMBH mass. We finally note that J0920's size would still correspond to a time lag of ~1200 days in the observer's frame, making reverberation mapping measurements more difficult and significantly longer compared to the few hours needed with GRAVITY+.

Our kinematic modelling infers a SMBH mass of log $M_{BH}$ = $8.51^{+0.27}_{-0.28}$ $M_\odot$, which we can compare to mass measurements using the "single-epoch" method from three different emission lines: CIV, Hβ, and Hα. Based on the CIV line width, we determine a mass of log $M_{BH}$ ~ 9.7 $M_\odot$, or about 1.2 dex larger than our spectro-interferometric result. Comparing the line profiles of CIV and Hα reveals that CIV is both systematically blueshifted by 5000 km s$^{-1}$ and significantly broader (full-width-at-half-maximum; FWHM ~ 8000 km s$^{-1}$ for CIV compared to 2500 km s$^{-1}$ for Hα). For J0920, CIV therefore must be tracing a high-velocity quasar-driven outflow rather than gravitationally bound gas, which reinforces concerns about adopting CIV-based single-epoch masses[23,24,25,26].

We determine a single-epoch Hβ mass of log $M_{BH}$ = 9.24 ± 0.47 $M_\odot$, which is 0.73 dex higher than our measurement from GRAVITY+ data. 0.53 dex of the discrepancy originates in the smaller BLR radius compared to that expected from the local radius-luminosity relation. The remaining discrepancy can be attributed to the *f* scaling factor needed to convert the single-epoch virial product to a black hole mass. This scaling factor has significant systematic uncertainty for individual objects as it is calibrated as a mean value such that a sample of AGN match the local $M_{BH}$-$\sigma_*$ relationship. The single-epoch Hα mass (log $M_{BH}$ =

8.94 ± 0.48 $M_\odot$) is only 0.43 dex larger, again due to the smaller BLR radius. While the single-epoch and spectro-interferometric Hα mass are reasonably in agreement, our GRAVITY+ based mass has much lower uncertainty given the ability to self-consistently measure size and mass and not rely on a scaling factor. Finally, we use the formalism of ref. 27 to correct the single-epoch Hβ BLR radius for the Eddington ratio and arrive at a BLR radius of 0.2 pc (Methods) and SMBH mass of 8.6 dex, now only 0.1 dex larger than the GRAVITY+ based mass and well within the uncertainties. Consequently, our spectro-interferometric result lends support to the idea that Eddington ratio is a nuisance factor in the radius-luminosity relation and that the correction proposed in ref. 42 may significantly improve single epoch mass estimates, especially for high luminosity quasars.

To investigate the host galaxy properties, we observed the CO (3-2) emission line for J0920 with the *NOEMA* interferometer which traces the molecular gas in the host galaxy and provides a measure of the galaxy mass even in the presence of the bright central quasar[28]. We infer a total dynamical mass, $\log(M_{dyn}/M_\odot) = 11.77^{+0.44}_{-0.37}$, and convert to a stellar mass using the average dynamical-to-stellar mass ratio found in $z\sim2$ star-forming galaxies[29] resulting in $\log(M_{stellar}/M_\odot) = 11.39^{+0.45}_{-0.39}$.

In Figure 3, we show J0920 on the $M_{BH} - M_{stellar}$ plane for $z \sim 2$. The two panels of Figure 3 split our comparison samples based on bolometric luminosity with high luminosity ($L_{bol} > 10^{47}$ erg s$^{-1}$) quasars on the right and lower luminosity ones on the left. For lower luminosity quasars, we use a sample of $z = 1.5 - 2.5$ galaxies from ref. 30 (gray points; left panel) for which both $M_{BH}$ and $M_{stellar}$ have been measured. $M_{BH}$ values for this sample were determined through the single-epoch method using the Hα, Hβ, or MgII broad emission line. Despite its higher luminosity, J0920 sits within the population of this sample. For high luminosity quasars we use the WISSH survey[31] (yellow points; right panel) quasars with published CO line measurements to convert them to $M_{stellar}$ in a similar way as for J0920[32]. $M_{BH}$ values are either based on single epoch measurements with the Hβ line[33] or CIV line[34]. J0920 lies well below the WISSH quasars, with a SMBH mass ~100 times smaller despite a comparable host galaxy mass and AGN luminosity. We point out that ~0.7 dex of the discrepancy can be alleviated if the deviation of the Hβ-based R-L relation at high luminosity or Eddington ratio holds true. In addition, the CIV-based masses may be significantly overestimated if, as for J0920, outflowing gas dominates the CIV line width. However, this only applies to half of the WISSH quasars. Even with these corrections, J0920 seems to have an undermassive SMBH given its luminosity and stellar mass that is more in line with more moderate luminosity quasars.

We further compare J0920 to the $M_{BH} - M_{stellar}$ local scaling relations, using a recent measurement of the relations for early (red line, Figure 3) and late type galaxies (blue line, Figure 3)[35]. J0920 lies firmly on the late type galaxy relation and well below the early type galaxy relation, consistent with a recent study of thousands of local AGN which found undermassive SMBHs typically have high accretion rates[36]. Massive, gas-rich galaxies at $z\sim2$ are thought to be the progenitors of massive ellipticals in the local universe[37]. These objects should therefore evolve onto the early type relation in Figure 3 by $z = 0$. J0920 would require more than a factor of ten growth in black hole mass and little growth in host galaxy stellar mass to reach this relation. The SMBH however is currently accreting material at an exceptionally fast rate of 30-140 $M_\odot$ yr$^{-1}$ depending on the specific bolometric correction (see Methods). Using an accretion rate of 85 $M_\odot$ yr$^{-1}$ we show as a blue arrow in Figure 3 the position of J0920 after $10^7$ years, which corresponds to the expected quasar lifetime[38]. J0920 would evolve directly onto the local early type galaxy relation. However, it is highly unlikely that the SMBH in J0920 would continue accreting material at such high super-Eddington rates for such a long time. Rather, several longer (~$10^8$ years) quasar episodes at more moderate Eddington ratios would be required to reach the local early type relation.

Some large-scale cosmological simulations predict that galaxies in the early universe out grow their SMBHs and attribute it to black hole growth in lower mass galaxies being inefficient[39,40]. One reason for this may be strong supernovae feedback where gas is quickly expelled from the central regions before it can reach the SMBH, and only when galaxies become massive enough to retain a nuclear gas reservoir against supernovae feedback do SMBHs begin to rapidly grow. This seems to be the likely scenario driving the evolution of J0920 given its current observed black hole mass, stellar mass, and black hole accretion rate. Whether this is the dominant mode of SMBH-galaxy co-evolution will only be revealed with more high-precision SMBH mass measurements.

## Main Figure Legends

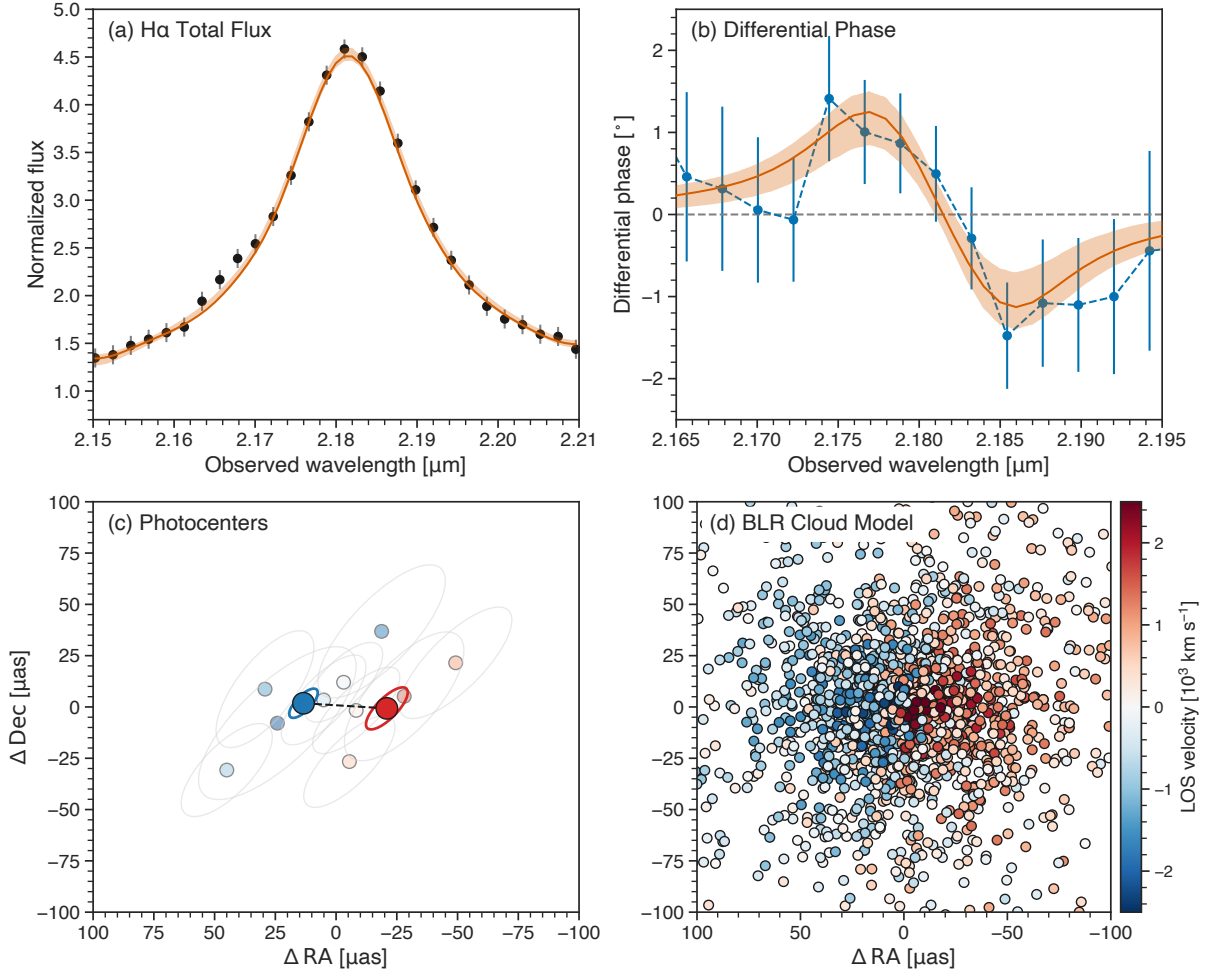

**Figure 1 | Main broad line region observational and modelling results. (a)** Observed GRAVITY+ Hα total flux line profile averaged over the four Unit Telescopes and normalized to the continuum (black points) with 1σ error bars. The red curve and shaded region indicate the line profile for our best fit BLR model and 68th %tile confidence region respectively. **(b)** Differential phase curve across the Hα line averaged over three baselines (blue points) with 1σ uncertainties. The red curve and shaded region also show the differential phase for our best fit BLR model and 68th %tile confidence region. The distinct S-shape signal is expected for a velocity gradient. **(c)** Model-independent photocenters for the central 10 wavelength channels (small coloured points). The colour of the points represents the line-of-sight velocity and the grey ellipses show the 68th %tile confidence region. The larger blue and red points with ellipses show the average blueshifted and redshifted photocenters with their 68th %tile confidence regions. **(d)** On-sky cloud representation of our

best fit BLR model showing an inclined, rotating, thick disk. As in (c), the colour represents line-of-sight velocity.

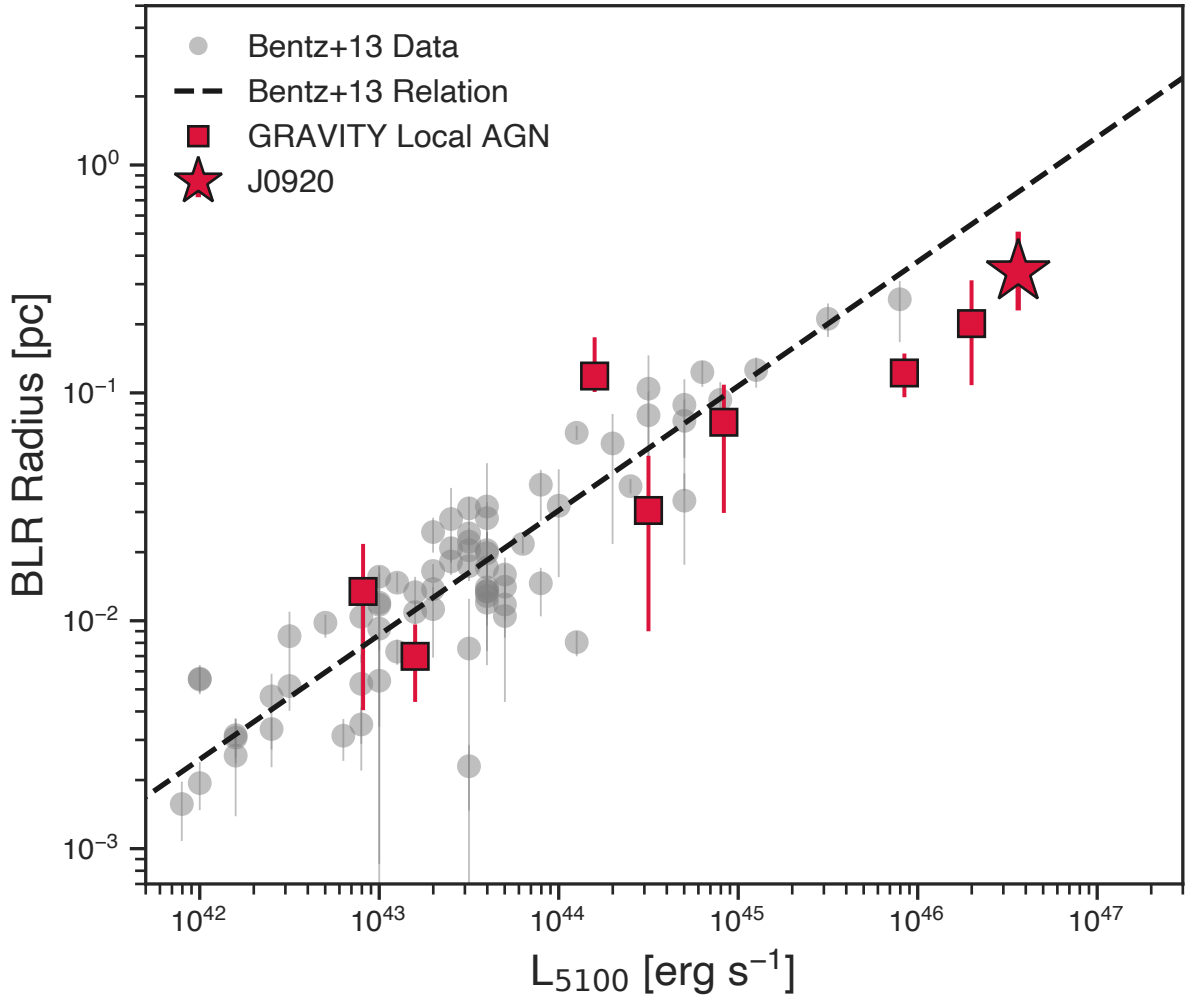

**Figure 2 | BLR Radius-Luminosity relation.** Empirical correlation between BLR radius and AGN luminosity (as measured by the luminosity at 5100 Angstrom). Gray points are reverberation mapping measurements from ref. 20. Moderate luminosity, local AGN measured by GRAVITY (red squares) [12,13,14] confirm the reverberation mapping based relation (ref. 11; dashed line). High luminosity quasars, including J0920 (red star), indicate a potential deviation from the relation towards smaller radii. All error bars represent $1\sigma$ uncertainties.

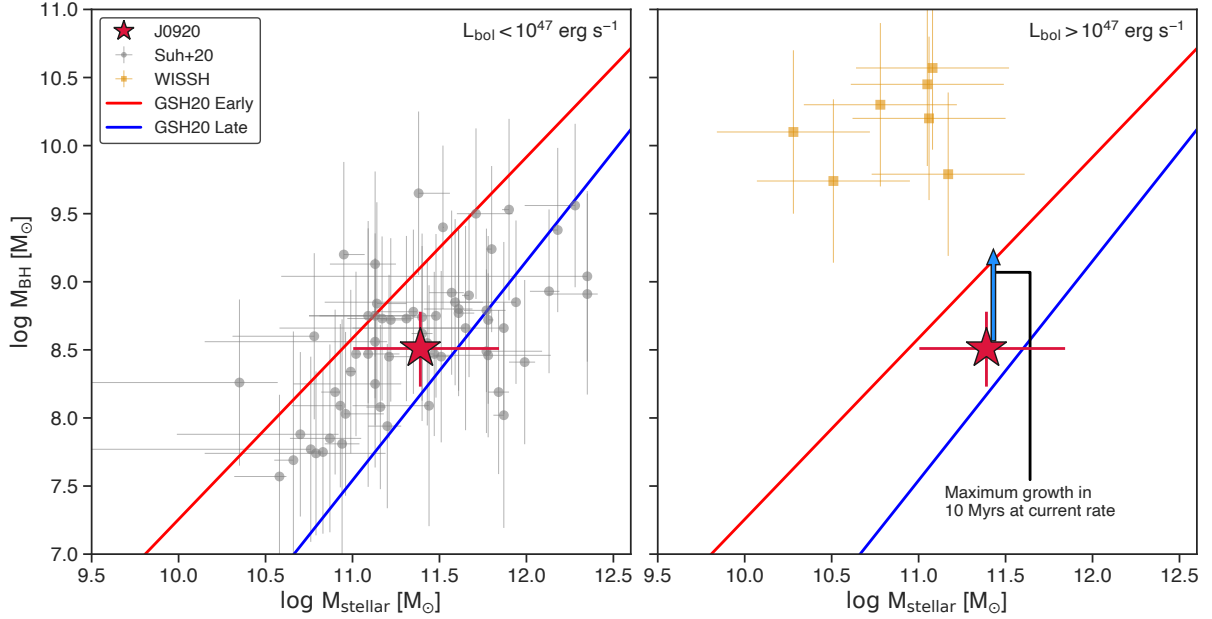

**Figure 3 | Black hole and host galaxy relation.** The location of J0920 in the SMBH mass – stellar mass plane (red star) compared to previously measured *z~2* AGN from ref. 15 (gray points) and the WISSH survey[52] (yellow squares). We split the figure into two panels based on the bolometric luminosity of the comparison sample with a cut at $L_{bol} = 10^{47}$ erg s$^{-1}$. Effectively this places all of the ref. 15 quasars in the left panel with lower luminosities and all of the WISSH quasars in the right panel with high luminosities. While J0920 has $L_{bol} > 10^{47}$ erg s$^{-1}$, we still plot it in both panels for comparison. GRAVITY+ provides a significantly improved constraint on the SMBH mass. J0920 clearly lies well below the high luminosity WISSH quasars and within the population of the ref. 15 sample showing the unique nature of J0920. Compared to recent local scaling relations[54]; J0920 is off the early type galaxy relation (red line) and near the late type galaxy relation (blue line). Given J0920's SMBH accretion rate, it should shift directly up towards the early type relation (blue arrow in right panel) and indicates it is currently in a state of rapid SMBH growth. All error bars represent 1σ uncertainties.

# Methods

### Target Selection

We selected J0920 from the *Million Quasar Catalog*[41] after associating each quasar to the nearest stars from the 2MASS Point Source Catalog. J0920 itself is detected in the 2MASS Point Source Catalog with a *K*-band Vega magnitude of 15.1 and is located 12.7 arcseconds away from the *K*=10.4 star, 2MASS 09203423+0657053. The initial redshift for J0920 (*z* = 2.30) was measured as part of the *LAMOST* Quasar Survey[42].

### GRAVITY+ Observations and Data Reduction

We observed J0920 at the VLTI with GRAVITY+ in the new GRAVITY-Wide mode as part of an Open Time Service Mode programme (PID: 110.2427, PI: T. Shimizu). We used the medium resolution (R~500) grating of the science channel spectrograph with combined polarization and the 300 Hz fringe tracking frequency. As the fringe tracking object, we used the star 2MASS 09203423+0657053. Science exposures consisted of four 100s detector integrations (DIT=100s, NDIT=4). A normal observing block (OB) was a sequence of six science exposures followed by a sky exposure where the science and fringe tracking fibers were moved 2" in right ascension and declination away from their nominal position.

OBs were executed over four nights on 2022-12-09, 2023-01-06, 2023-01-10, and 2023-01-11 under excellent weather conditions (average seeing = 0.48", average coherence time = 11.3 ms). We obtained in total 32 exposures (128 DITs) resulting in an on-source integration time of 3.56 hours. However, on 2023-01-06, the UT4 science channel fiber was positioned off the quasar. Therefore, only the three non-UT4 baselines from this night are used for further analysis.

We first used the standard GRAVITY pipeline[43] (v1.4.2) to reduce all raw files up to the application of the pixel-to-visibility matrix (P2VM). This means the pipeline performed the bias and sky subtraction, flat fielding, wavelength calibration, and spectral extraction steps. Application of the P2VM converts the pixel detector counts into complex visibilities taking into account all instrumental effects including relative throughput, coherence, phase-shift, and cross-talk. This results in four complex visibility spectra per baseline per exposure covering the 1.97 – 2.48 μm wavelength range.

At this point, we proceeded to process the intermediate products (i.e. *dualscip2vmred.fits* files) with our own scripts. This was meant to mitigate potential effects related to the unique situation where the majority of the signal is within the emission line and not the continuum. We first measured the coherent flux within the line by summing the spectral channels between 2.17 and 2.19 μm, covering roughly the FWHM of the line. We removed frames where the integrated emission line coherent flux was less than $10^{3.5}$ counts. This limit was chosen based on the integrated emission line coherent flux measured on the UT4 baselines from 2023-01-06. On this night, the science channel fiber for UT4 was not positioned on the quasar so any measured coherent flux is noise. Frames showed a maximum emission line coherent flux of $10^{3.5}$ counts which we then chose as our threshold for accepting frames on other nights. For the selected frames, we first subtracted the pipeline measured self-referenced phases which are a third-degree polynomial fit to the whole wavelength range of each visibility spectrum. We then cut out the 2.10 – 2.26 μm region and measured and subtracted a second third-degree polynomial to the visibility phases to remove any remaining residual instrumental phase and produce the differential phase spectra. To avoid large outliers influencing the fit, we used the *FittingWithOutlierRemoval* function within the *astropy.modeling* module[44] to iteratively perform fits and at each step remove all channels more than 3σ away from the previous best fit. The stopping criterion is then when either no channels are thrown away or five iterations is reached. On average only 1-2 iterations were needed per baseline. Finally, we averaged over time all phase-flattened complex visibilities per baseline and calculated the resulting average differential phase spectra. Phase uncertainties per spectral channel were measured with the method described in ref 45. At high signal-to-noise, this simply reduces to the standard error of the mean. The averaged differential phase spectra through the inner part of the Hα line are shown in Extended Data Figure 1.

To calibrate the total flux spectrum, we used the data from 2022-12-09 where the OBs were executed directly after the observation of a bright binary star pair calibrator with GRAVITY-Wide. We reduced the calibrator data using the same pipeline and divided J0920's spectra by the calibrator spectra for each telescope to remove the atmospheric and instrumental response. We then averaged together the four spectra to produce a single total flux spectrum for J0920. As the differential phase and BLR modelling is only sensitive to the line-to-continuum ratio, we measured the underlying continuum by fitting a second-degree polynomial to the 2.05 – 2.10 and 2.25 – 2.35 μm regions. The best fit continuum was then divided out of the flux spectrum for the final normalized line profile. The line profile is shown in Extended Data Figure 1 and Figure 1a of the main text. As uncertainty on the line profile, we measure the root-mean-square variation in the continuum fitted regions finding 0.05. We multiply this by a factor of 2 to conservatively account for systematic effects.

**Photocenter Measurement**

The first analysis performed on the GRAVITY differential phases and line profile is the measurement of model-independent photocenters as a function of wavelength/velocity. We use the same procedure as in previous AGN observations[12,13,14] and briefly describe it here. In the marginally resolved limit, the differential phase, $\Delta\Phi_{ij} = -2\pi f_{line}(u_j x_i + v_j y_i)$ where $i$ runs across wavelength and $j$ runs across baselines. $(u_j, v_j)$ are the projected baseline coordinates and $(x_i, y_i)$ the on-sky photocenter coordinates for each spectral channel. $f_{line} = f_i / (1 + f_i)$ where $f_i$ is the line intensity as a fraction of the continuum. We use the *emcee* package[46] to perform Markov Chain Monte Carlo sampling to fit for $(x_i, y_i)$ of the central 10 spectral channels across the Hα line and sample the posterior. We use the median of each marginalized posterior as our best photocenter positions and determine the uncertainty by fitting a 2D Gaussian to the joint posterior of each $(x_i, y_i)$ pair. The best fit photocenters and uncertainties are shown in Figure 1c where we clearly see red and blueshifted positions on opposite sides of the central channel along a line in the East-West direction.

We also measure an average red-blue offset which we term the "2-pole" model. To do this, we first set the central wavelength (2.182 μm) to define which channels are redshifted and which are blueshifted. The model then assumes all redshifted channels share the same photocenter coordinate ($x_{red}$, $y_{red}$) and all blueshifted channels share the same photocenter coordinate ($x_{blue}$, $y_{blue}$). We further include a systematic shift of the BLR shared by all channels, ($x_{off}$, $y_{off}$). The fitting is performed in the same way as above, just with only two photocenter coordinate pairs as the free parameters. We find ($x_{blue}$, $y_{blue}$) = (13.6, 1.6) +/- (5.8, 7.0) uas and ($x_{red}$, $y_{red}$) = (-20.6, -0.6) +/- (8.6, 10.1) uas, which are shown as the large points in Figure 1c. The $\chi^2_{\text{2-pole}} = 38.8$.

Finally, we perform a third fit now assuming all spectral channels lie at the same photocenter ($x_{null}$, $y_{null}$) and the BLR is completely unresolved. This results in either differential phase spectra equal to 0 at all wavelengths (if $x_{null} = y_{null} = 0$) or differential phase spectra with the same shape as the emission line profile. We find ($x_{null}$, $y_{null}$) = (3.3, -3.6) +/- (3.8, 9.8) uas with $\chi^2_{null} = 54.3$.

We use an *F*-test to compare the "2-pole" and null model and determine whether the "2-pole" model gives a significantly better fit. The *F* statistic is $\frac{\left(\frac{\chi^2_{null} - \chi^2_{\text{2-pole}}}{p_{\text{2-pole}} - p_{null}}\right)}{\left(\frac{\chi^2_{\text{2-pole}}}{n - p_{\text{2-pole}}}\right)}$ where the $\chi^2$ are the total $\chi^2$ from each fit, $p$ are the number of parameters for each model, and $n$ is the number of data points used in the fit. We calculate $F = 5.41$, which corresponds to a *p*-value of $10^{-9}$ and a significance of 6σ to reject the null model.

To test for systematics, we downloaded 22 archival calibrator observations in the GRAVITY-Wide mode which results in 664 individual frames that have signal-to-noise comparable to J0920. These data should have zero differential phase since they are single stars, and therefore allow for testing while including systematics. We processed the calibrator data in the same manner as J0920 and measured the average redshifted and blueshifted positions using the same wavelength channels and emission line profile. We fit the distribution of red-blue separations with a truncated Gaussian finding a standard deviation of 12 uas. Given the measured separation for J0920 of 37 uas, this indicates a significance of at least 3σ. We consider this a lower limit because we did not specifically test how often the broader S-shape signal of J0920 occurs. Rather it is likely that many of the non-zero red-blue separations measured in the calibrator data are caused by narrow noise spikes.

## BLR modeling

Our primary analysis centers on modelling the BLR structure and kinematics using the GRAVITY observed differential phase and total flux spectra. We refrain from a detailed description of the model and fitting procedure as this has been outlined in several past publications[12,13,14]. In general, we model the BLR as a set of independent, non-collisional clouds solely under the gravitational influence of the central SMBH. The model very closely follows the one used to fit reverberation mapping data[47,48] with the major adjustment to output differential phases instead of light curves[49]. While the model contains a number of parameters to introduce deviations away from the axisymmetric Keplerian model, we choose to omit those and only use the minimal number of parameters able to best describe our data. The fitted model therefore contains eleven free parameters: $R_{BLR}$, $\beta$, PA, $\theta_0$, $i$, $F$, $M_{BH}$, $f_{peak}$, $\lambda_{emit}$, $x_0$, and $y_0$. A brief description of each parameter along with the prior distributions used in the fitting is given in Extended Data Table 1.

We fit the model to both the total flux spectrum and six baseline averaged differential spectra. We fit only the central 2.15 – 2.21 μm region with the highest signal-to-noise-ratio but note that fits over the entire 2.1 – 2.26 wavelength range do not produce significantly different results. We used the *dynesty* package[50] (v2.1) which performs Dynamic Nested Sampling[51] to sample the potentially complicated posterior. We used multi-ellipsoidal decomposition to bound the target posterior distribution (*bound = 'multi'*) and the random walk sampling method. Sampling was done with 2000 live points and we chose to stop sampling once the iterative change in the logarithm of the evidence is less than 0.01 (*dlogz_init = 0.01*).

In Extended Data Figure 2, we plot the 2D joint and 1D marginalized posterior distributions. The posteriors are well sampled and largely show symmetric, Gaussian shaped posteriors. We report in Extended Data Table 1 the medians of each 1D marginalized posterior distribution and as uncertainties the 68$^{th}$ percentile confidence interval. We further plot the prior distributions for each parameter used in the modelling with the 1D marginalized posterior distributions. The posteriors have significantly shifted and/or narrowed from the initial prior showing the data well constrains each parameter.

To test for potential systematic errors, we fit the data with the full kinematic model including all asymmetric parameters and radial motion. Even though this adds another seven extra free parameters, the reduced chi-square is not improved compared to the simpler axisymmetric model and the posteriors of the extra parameters largely indicate they are unconstrained with distributions similar to the input priors. An advantage of *dynesty* is the measurement of the Bayesian evidence ($Z$) which can be used to compare models. We find $\ln(Z_{sym})$ = -333 for the axisymmetric model and $\ln(Z_{full})$ = -332 for the full model. The ratio of the evidences, or Bayes factor, then quantifies the support for one model over the other. We calculate a Bayes factor, $Z_{full}/Z_{sym}$ = 2.7 which indicates weak support for the full model over the simpler, axisymmetric model. We further note that the uncertainties on all of the original parameters do not significantly increase. However, the median of the posterior for the SMBH mass does slightly increase from log $M_{BH}$ = 8.51 to 8.67. This shift is within the 1σ uncertainty but suggests an additional potential systematic uncertainty. We therefore add in quadrature 0.16 dex to the statistical uncertainty of the black hole mass resulting in a final uncertainty of 0.27 and 0.28 dex for the upper and lower uncertainties respectively.

## APO/TripleSpec Observations and Data Reduction

We observed J0920 with the TripleSpec instrument at *Apache Point Observatory* for 56 min on Dec. 21, 2021 with a slit width of 1.1" providing a spectral resolution of 3181 over the H and K wavelength bands.

## APO/TripleSpec Emission Line Measurements

The TripleSpec spectrum provides the rest-frame optical spectrum of J0920 at much higher spectral resolution compared to GRAVITY+ and covers the Hβ-[OIII] region. This provides an opportunity to compare our spatially resolved BLR size and dynamically measured SMBH mass with those inferred from the single epoch method. We first scaled the H-K band spectrum to match the K-band magnitude of J0920 from the 2MASS Point Source Catalog (K=15). We simultaneously fit the continuum, FeII features, Hα, Hβ, and [OIII] doublet and adopt a fourth-order polynomial to describe the continuum combined with the FeII template from ref 52. To model the [OIII] doublet we use a single Gaussian component while fixing the [OIII] doublet flux ratio to the theoretical value of 2.98[53] and tying the velocity and line width together for the two components of the doublet. While for Hβ we use only a single Gaussian component; for Hα, we found we needed two Gaussian components to adequately fit the line but note we do not consider each component to be tracing different physical components of the emission. Rather, the line profile is likely better described by a Lorentzian shape. We find very good agreement between the TripleSpec line profile and GRAVITY+ line profile after degrading the TripleSpec line profile to the spectral resolution of GRAVITY+ indicating we are not seeing extra, more extended narrow line emission in the much larger aperture of TripleSpec. In Extended Data Table 2, we list the best fit parameters of our spectral decomposition as well as the derived properties and show in Extended Data Figure 3 the best fit model and decomposition along with the residuals. The fitting residuals are about $10^{17}$ erg s$^{-1}$ cm$^{-2}$ Angstrom$^{-1}$ around 5000 Angstrom and 0.7 x $10^{17}$ erg s$^{-1}$ cm$^{-2}$ Angstrom$^{-1}$ around 6500 Angstrom. The uncertainties of the measured quantities are derived by refitting the spectra after adding Gaussian noise with a standard deviation equal to the fitting residual at the corresponding wavelength.

In the Table, EW is defined as the equivalent width. $R_{Fe}$ is defined as the ratio of FeII template equivalent width within 4434–4684 Angstrom to Hβ equivalent width. $L_{5100}$ is the monochromatic luminosity at rest-frame wavelength 5100 Angstrom. We first calculate the bolometric luminosity ($L_{Bol}$) using the empirical relation from ref 54 which is based on an average luminosity dependent quasar spectral energy distribution. The bolometric correction here is ~5 and already placing J0920 well into the super-Eddington regime. Therefore, we also estimate the bolometric luminosity under the slim disk accretion model which is theorized to be applicable for highly accreting black holes. We use Equation 3 from ref. 55 to determine a bolometric correction of ~23. The bolometric luminosities for both corrections are listed in Extended Data Table 2. From the bolometric luminosity, we estimate a mass accretion rate onto the SMBH of $\dot{M} = L_{Bol}/\eta c^2$ M☉ yr$^{-1}$ using a standard conversion efficiency, $\eta = 0.1$.

## Comparison to Single-Epoch Estimates

### CIV

Our first comparison is to the CIV based mass estimate which was measured for the LAMOST QSO catalog[42]. The reported redshift and FWHM of the CIV line are 2.3015 and 8013 km s$^{-1}$ respectively. They use the CIV radius luminosity relation from ref. 56 to determine a SMBH mass of $10^{9.7}$ M☉. Compared to our Hα measurements, the redshift is off by 0.0235 (7050 km s$^{-1}$), the FWHM is a factor ~3 larger, and the SMBH mass is 1.2 dex larger. In Extended Data Figure 4, we compare the line profiles of CIV and Hα using $z = 2.325$ to convert wavelengths into velocities. This shows clearly the significant blueshift of the CIV line relative to the systemic velocity of Hα as well as the increased line width. Since single epoch masses scale with FWHM$^2$, the factor of 3 larger FWHM mostly explains the factor of 15 increase in the SMBH mass. Beyond the systematic blueshift of CIV, the line shape is also heavily skewed towards large blueshifted velocities. All of these properties point to CIV emission being dominated by non-virial motions and likely originating in a strong outflow[57,58]. Previous surveys of high redshift quasars have reported strong correlations between CIV blueshift and FWHM and an anti-correlation between CIV blueshift and Hα FWHM[59,60] which leads to CIV overestimating the SMBH mass. In fact, ref. 59 provides a correction to CIV based masses based on the blueshift and FWHM of CIV. Applying this (see Equations 4 and 6 of ref. 59) to J0920, we calculate a corrected CIV SMBH mass of $10^{8.7}$ M☉, much closer to our dynamically based mass.

### Hα and Hβ

We further compare our GRAVITY based BLR size and SMBH mass with the single epoch sizes and masses inferred from the Hα and Hβ relations. We first calculate $R_{BLR}$ from an extrapolation of the "Clean2" Hβ radius-luminosity relation from ref. 11: log $R_{BLR}$ = 1.56 + 0.546 log($L_{5100}$ / 10$^{44}$ erg s$^{-1}$) [light-days]. This gives $R_{BLR}$ = 907 light-days or 0.765 pc which is a factor of 2.25 times larger than our spatially resolved measurement. This R-L relation has a scatter of 0.13 dex so our smaller size is 1.65σ away from the best fit. If the Hα emitting region is larger than the Hβ emitting region as observationally found from reverberation mapping studies[56,16,17] and expected from BLR photoionization models[18], then our BLR size is even more discrepant from the radius-luminosity relation size.

We estimate the Hβ single epoch SMBH mass using the standard virial relation: $M_{BH} = f(R_{BLR}\Delta v^2/G)$ where $\Delta v$ is a measure of the line width and $f$ is a scale factor that accounts for the orientation and geometry of the BLR. For $\Delta V$, we choose to use the Hβ FWHM. We further use log <$f$> = 0.05±0.12, which was determined empirically by fitting Hβ FWHM based black hole masses onto the local $M_{BH}$-$\sigma_*$ relation[61]. The intrinsic scatter associated with the Hβ single epoch calibration is measured to be 0.43 dex[56]. The Hβ single-epoch black hole mass then is log $M_{BH}$ = 9.24 ± 0.47, which is 0.73 dex larger than our dynamical measurement. Taking into account the expected factor of 1.5 larger sizes for the Hα emitting region[16], then 0.53 dex of the discrepancy can be explained by the significantly smaller BLR we measure with GRAVITY+. The remaining 0.2 dex can then be explained by scatter in BLR inclination and geometry leading to variations in individual $f$ scale factors.

We use Equation 1 from ref 62 to calculate the H$\alpha$ single epoch mass, which was calibrated off the H$\beta$ R-L relation and a correlation between the FWHM of H$\beta$ and H$\alpha$ and between $L_{5100}$ and $L_{H\alpha}$: $\log(M_{BH}/M_\odot) = \log(f) + 6.57 + 0.47 \log(L_{H\alpha}/10^{42} \text{ erg s}^{-1}) + 2.06 \log(FWHM_{H\alpha}/1000 \text{ km s}^{-1})$. Using the same $f$ scaling factor as before, we find $\log(M_{BH}/M_\odot) = 8.94 \pm 0.48$ which is only 0.43 dex larger than our dynamical measurement and within the uncertainties of the single-epoch measurement. This can then be fully explained by the smaller BLR size we measure compared to the expectation from the R-L relation.

Deviations from the standard R-L relation have been seen and explored before with most of the scatter leading to smaller sizes for a given AGN luminosity[20,21,22,63,55]. Ref. 55 found that the offset from the R-L relation was correlated with the Eddington ratio. After gathering a large sample of reverberation mapping measurements for high-Eddington ratio targets through the Super-Eddington Accreting Massive Black Hole (SEAMBH) survey, ref. 27 proposed a new parameterization of the R-L relation including $R_{Fe}$, the flux ratio of Fe II features between 4434-4684 Angstrom and broad H$\beta$. Eddington ratio has been shown to be the dominant property driving variations in $R_{Fe}$ between AGN[64,65,66] and so including $R_{Fe}$ implicitly adds a second property determining the BLR size beyond the AGN luminosity. The new parameterization is $\log R_{BLR} = 1.65 + 0.45 \log(L_{5100}/10^{44} \text{ erg s}^{-1}) - 0.35 R_{Fe}$ [light-days]. With this we calculate an Eddington ratio corrected BLR size of 237 light-days or 0.2 pc, a factor of 1.7 *smaller* than our GRAVITY measured size. This then leads to a $\log(M_{BH}/M_\odot) = 8.66$ using the same $f$ scaling factor as above, the closest "single epoch" estimate to our dynamical measurement. While J0920 is only one object, this certainly adds to the evidence that the BLR size is related to the Eddington ratio of the SMBH and thus should be taken into account for SMBH mass measurements.

## NOEMA Observations and Data Reduction

To complement our GRAVITY observations and probe the host galaxy of J0920, we observed J0920 with the IRAM NOrthern Extended Millimeter Array (NOEMA) as part of a larger pilot survey of $z \sim 2$ quasars (ID: S22CE, PI: J. Shangguan) on 12 June and 18 September 2022 in D configuration. The total on-source time was 3.9h with 10 antennae. We set the phase center to the known coordinates of J0920 (RA = 09:20:34.171, DEC = 06:57:18.019) and used the PolyFix correlator with a total bandwidth of 15.5 GHz. With a tuning frequency of 104.7867 GHz, we placed the redshifted CO (3-2) molecular gas line ($\nu_{rest}$ = 345.7960 GHz, $\nu_{obs}$ = 103.99 GHz) into the upper sideband.

The sources J0923+392, J2010+723, J0906+015, and J0851+202 were used as flux calibrators and J0906+015 and J0851+202 were used for phase calibration. Observations were taken under average weather conditions with precipitable water vapor of 4-10 mm. We reduced and calibrated the data with the *CLIC* package of *GILDAS* to produce the final (u,v) tables.

The (u,v) tables were then imaged with the *MAPPING* package of *GILDAS* using the *hogbom* CLEAN algorithm. We adopted natural weighting of the visibilities resulting in a synthesized beam of 4.7"x3.2". We ran CLEAN until the maximum of the absolute value of the residual map was lower than 0.5$\sigma$ with $\sigma$ the RMS noise of the cleaned image and used a circular support mask with diameter 18" centred on J0920. We then re-sampled the spectral axis to 40 km s$^{-1}$ bins achieving an RMS noise of 0.388 mJy beam$^{-1}$.

## Host Galaxy Properties

In the top left panel of Extended Data Figure 5, we show the 0$^{th}$ moment image of the cube generated between -700 and 700 km s$^{-1}$ around the expected location of the CO (3-2) line. We clearly detect J0920 with a maximum SNR of >20 and visual comparison of the image with the synthesized beam suggests J0920 is extended especially in the North-South direction. To test this and measure a CO size, we used UVFIT within *GILDAS* to fit the visibilities directly with an elliptical exponential disk model. The top right panel of Extended Data Figure 5 shows the visibilities as a function of baseline length together with our best fit model. The clear drop with baseline length is indicative of a partially resolved source. Our best fit disk model fits the data well and confirms the resolved nature. A Gaussian disk model provides a nearly equally good fit and the same effective radius as the exponential disk considering the uncertainty. We prefer to adopt the results with the exponential disk model to facilitate estimating the dynamical mass of the host galaxy using the empirical relation of ref 28.

We measure an effective radius of the disk, $R_e$ = 8.23 +/- 1.53 kpc, a position angle on sky of 90.0° +/- 0.4°, and an axis ratio of 1.66 +/- 0.8. This places J0920 at the upper envelope of the size-mass relation for its redshift and firmly within the late-type galaxy population[67] under the assumption that the molecular gas disk traces the stellar disk.

To measure the CO (3-2) flux and linewidth, we extracted a 1D spectrum by integrating the cube within the 1σ contour of the 0$^{th}$ moment map. We plot the resulting spectrum in the bottom panel of Extended Data Figure 5 which shows clearly the CO (3-2) line. We fit the line from the integrated spectrum with a single Gaussian component finding a redshift of 2.3253 +/- 0.0002 (very similar to the Hα redshift), integrated flux of 2.330 +/- 0.162 Jy km s$^{-1}$, and a FWHM of 432 +/- 42 km s$^{-1}$.

Ref. 28 provides empirical relations between the dynamical mass of a system and unresolved, integrated line properties based on spatially resolved kinematic modeling of z~ 6 quasar host galaxies. Here we use Equation (15) which assumes robust measurements of the line FWHM and radial extent of the galaxy have been made, as in the case of J0920: $M_{\rm dyn} = 1.9^{+1.5}_{-0.8}(^{+1.1}_{-1.3}) \times 10^5 ({\rm FWHM})^2 R_e$ [M$_\odot$] where FWHM is in km s$^{-1}$ and $R_e$ is in kpc. For J0920, we find $log(M_{\rm dyn}/M_\odot) = 11.77^{+0.44}_{-0.37}$ where the uncertainties are a combination of the measurement errors of the line and the statistical (first set of uncertainties in equation) and systematic uncertainties (second set of uncertainties in equation) of the empirical relation.

To infer the stellar mass, we use the empirically determined average dynamical-mass-to-stellar-mass ratio for $z$ = 2.0 – 2.6 galaxies from ref. 29, log(M$_{\rm dyn}$/M$_{\rm stellar}$) = $-0.38^{+0.11}_{-0.11}$. This results in a stellar mass of $log(M_{\rm star}/M_\odot) = 11.39^{+0.52}_{-0.48}$ M☉.

As a check on the stellar mass, we also convert the integrated CO (3-2) flux into a CO line luminosity, $L'_{CO}$, using the standard formula from ref 68: $L'_{\rm CO} = 3.25 \times 10^7 \frac{S_{\rm CO} R_{13} D_L^2}{(1+z)\nu_{\rm rest}^2}$ K km s$^{-1}$ pc$^2$ where $S_{\rm CO}$ is the CO line flux in Jy km s$^{-1}$, $D_L$ is the luminosity distance in Mpc, $z$ is the redshift, and ν$_{\rm rest}$ is the rest frequency of the line in GHz . $R_{13}$ is the CO (1-0)/CO (3-2) brightness temperature ratio such that $L_{\rm CO}$ is referred to the CO (1-0) line. We adopt $R_{13}$ =

0.97, a typical value for quasars[69] with which we find $L'_{CO}$ = 6.91 x $10^{10}$ K km $s^{-1}$ $pc^2$. We then convert this to a molecular gas mass using the CO-$H_2$ conversion factor, $\alpha_{CO}$, which we take as 4.36 $M_\odot$ [K km $s^{-1}$ $pc^2$]$^{-1}$ [70, 71, 72] with a 30% uncertainty. This results in a total molecular gas mass of log ($M_{H2}/M_\odot$) = 11.48 +/- 0.13.

Combining the molecular gas mass and stellar mass leads to a molecular gas fraction of 0.55, consistent with gas fractions of massive star-forming galaxies at z~2[71]. The baryonic fraction, ($M_{stellar}$ + $M_{H2}$)/$M_{dyn}$ is then 0.93 indicating little dark matter within the effective radius of the host galaxy, also consistent with deep, spatially resolved observations of $z$ = 2 star-forming galaxies[73,74,75]. Therefore, if we would have made the assumption that the dynamical mass is entirely composed of the stellar and molecular gas mass, we would have arrived at log($M_{stellar}$) = 11.45, completely consistent with the stellar mass derived from the dynamical-to-stellar mass ratio.

# Methods References

## Data Availability

The GRAVITY+ data used in this study are publicly available on the ESO archive (https://archive.eso.org/eso/eso_archive_main.html) under Programme ID 110.2427. The NOEMA and APO/TripleSpec data are available from the corresponding author upon request.

## Code Availability

The GRAVITY data reduction pipeline is publicly available on the ESO webpage (https://www.eso.org/sci/software/pipelines/). *GILDAS* is publicly available on the IRAM webpage (https://www.iram.fr/IRAMFR/GILDAS/). *Astropy*, *Matplotlib*, *emcee*, *dynesty*, *numpy,* and *scipy* are all available through the Python Package Index (https://pypi.org). The custom photocenter fitting and BLR modelling packages are available upon request from the corresponding author.

## Acknowledgements

GRAVITY+ is developed by the Max Planck Institute for extraterrestrial Physics, the Institute National des Sciences de l'Univers du CNRS (INSU) with its institutes LESIA / Paris Observatory-PSL, IPAG / Grenoble Observatory, Lagrange / Côte d'Azur Observatory and CRAL / Lyon Observatory, the Max Planck Institute for Astronomy, the University of Cologne, the CENTRA - Centro de Astrofisica e Gravitação, the University of Southampton, the Katholieke Universiteit Leuven and the European Southern Observatory. We are very grateful to our funding agencies (MPG, DFG, BMBF, ERC, CNRS (CSAA, ASHRA), Ile-de-France region (DIM ACAV+), Paris Observatory-PSL, Observatoire des Sciences de l'Univers de Grenoble, Université Grenoble Alpes, Observatoire de la Côte d'Azur, Université Côte d'Azur, and the Fundação para a Ciência e Tecnologia) and the generous support from the Max Planck Foundation - an independent, non-profit organization of private supporters of top research in the Max Planck Society. We are also grateful to ESO and the Paranal staff, and to the many scientific and technical staff members in our institutions, who helped to make GRAVITY and GRAVITY+ a reality. F.W. has received funding from the European Union's Horizon 2020 research and innovation programme under grant agreement No 101004719. DD, MS, and RL acknowledge the support from the European Research Council (ERC) under the European Union's Horizon 2020 research and innovation programme (Grant Agreement No. 866070). J.S.-B. acknowledges the support received from the UNAM PAPIIT project IA 105023; and from the CONAHCyT "Ciencia de Frontera" project CF-2019/263975. The research leading to this work was supported by the French government through the ANR AGN_MELBa project (reference number ANR-21-CE31-0011) and by the European Union's Horizon 2020 Research and Innovation Programme under grant agreement No. 101004719 (OPTICON RadioNet Pilot).

## Author Contributions

T.S., J.S., R.D., E.S., S.F.H., T.P., F.M., J.D., F.E., D.J.S., T.d.Z., Y.C., N.M.F.S., J.L., and J.B.L.B. wrote the proposal and designed the programme for the GRAVITY+ observations. T.S. and J.S. processed, analysed, and modelled the GRAVITY+ data. J.S., Y.C., R.D., J.D.,

T.d.Z., N.M.F.S., R.G., D.L., D.L., D.J.S., T.S., E.S., and L.J.T. wrote the proposal and designed the programme for the NOEMA observations. J.S. and D.L. processed and analysed the NOEMA data. T.S., J.S., R.G., S.F.H., E.S., D.L., R.D., N.M.F.S., D.J.S., T.d.Z., R.P., J.L., J.B.L.B., A.S., and H.N. drafted the text and figures. All authors participated in either developing and building GRAVITY+, discussing the interpretation of the data, and/or commented on the manuscript.

## Author Information

The authors declare no competing financial interests. Correspondence and requests for materials should be addressed to T.S. (shimizu@mpe.mpg.de). Reprints and permissions information is available at www.nature.com/reprints.

## Extended Data Figure and Table Legends

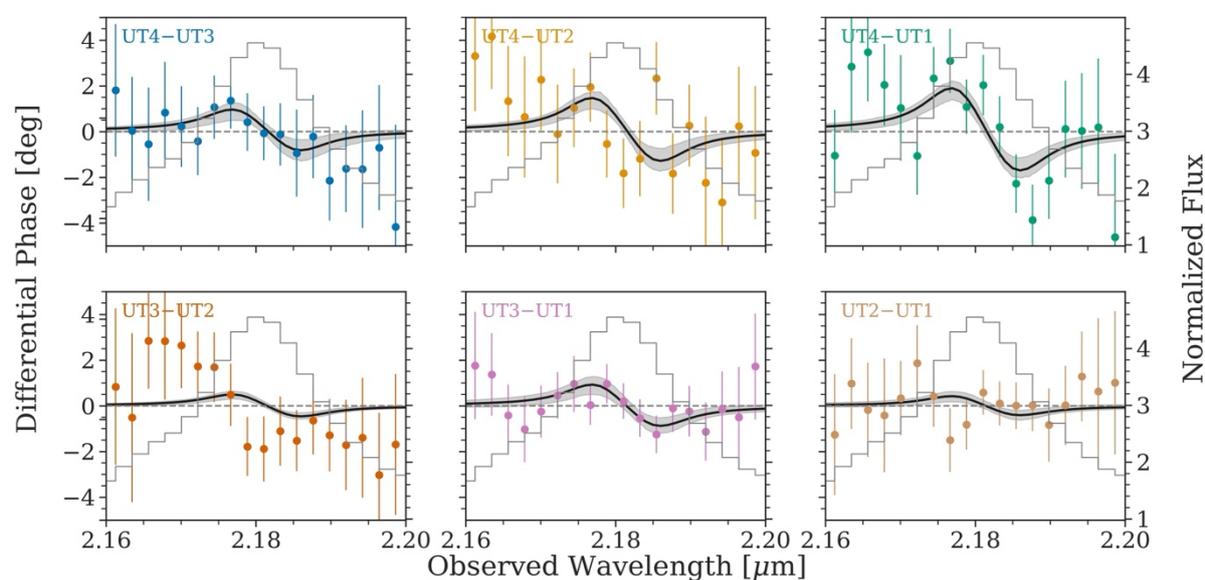

**Extended Data Figure 1 | Individual baseline differential phase spectra.** Average differential phase spectra for each baseline in the 2.16 – 2.20 μm region (colored points with 1σ error bars) together with the total flux spectrum (gray line) and best fit BLR model (black line) with 68[th]%tile confidence region (shaded region).

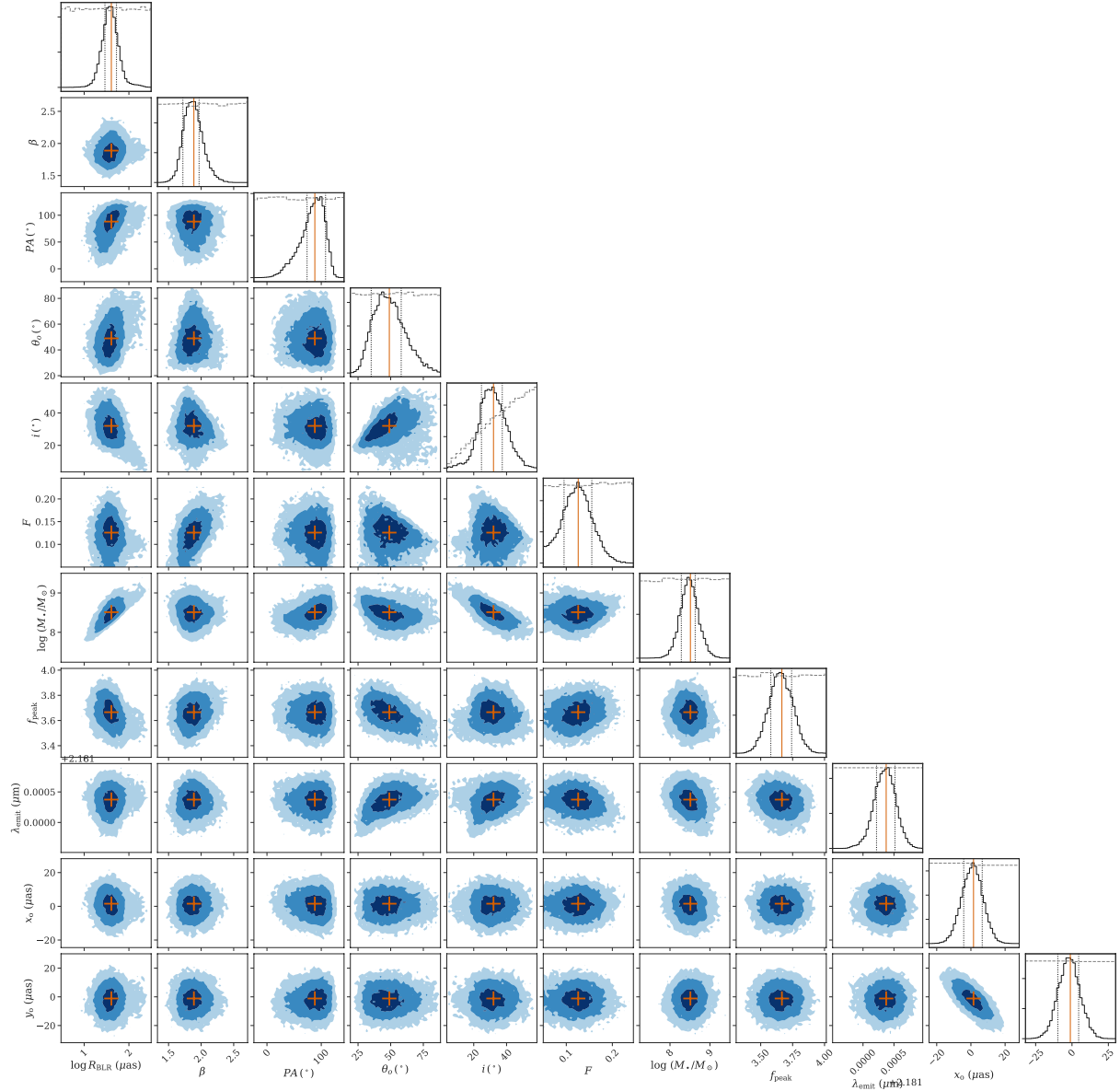

**Extended Data Figure 2 | Corner plot of the 2D and 1D posterior distributions for the BLR fit.** We plot the 2D joint and 1D marginalized posterior distribution for each parameter of the BLR model used to fit the differential phase and flux spectra. Blue shaded contours represent the 1, 2, and 3σ regions and the orange crosses are the median values that are also reported in Table 1. The dotted lines in the 1D posteriors indicate the 16th and 84th percentiles. The orange lines are again the median value. The dashed gray lines show a sampling of the priors used in the fitting which are listed in Extended Data Table 1.

| PARAMETER | VALUE | PRIOR | DESCRIPTION |
|---|---|---|---|
| $R_{BLR}$ | $40^{+20}_{-13}$ | LogUniform(3, 380) | Mean radius (μas) |
| β | $1.9^{+0.14}_{-0.13}$ | Uniform(0, 3) | Radial distribution shape |
| P.A. | $87°^{+19}_{-25}$ | Uniform(0, 2π) | Position angle East of North (deg) |
| $θ_0$ | $49^{+13}_{-11}$ | Uniform(0, 90) | Angular thickness of disk (deg) |
| i | $32^{+8}_{-7}$ | CosUniform(0, 60) | Inclination angle (deg) |
| F | $0.12^{+0.03}_{-0.04}$ | Uniform(0.05, 0.4) | Ratio of minimum to mean radius |
| log $M_{BH}$ | $8.51^{+0.22+0.16}_{-0.23-0.16}$ | Uniform(6, 12) | Black hole mass including systematic uncertainty |
| $f_{peak}$ | $3.7^{+0.1}_{-0.1}$ | Uniform(3, 6) | Amplitude of line profile |
| $\lambda_{EMIT}$ | $2.1814^{+0.0002}_{-0.0002}$ | Gaussian(2.182, 0.01) | Central wavelength (μm) |
| $(x_0, y_0)$ | $(1^{+6}_{-6}, -1^{+7}_{-7})$ | Uniform($-10^3$, $10^3$) | Systematic shift of BLR relative to the continuum (μas) |

**Extended Data Table 1 | BLR model parameters and fit values.**

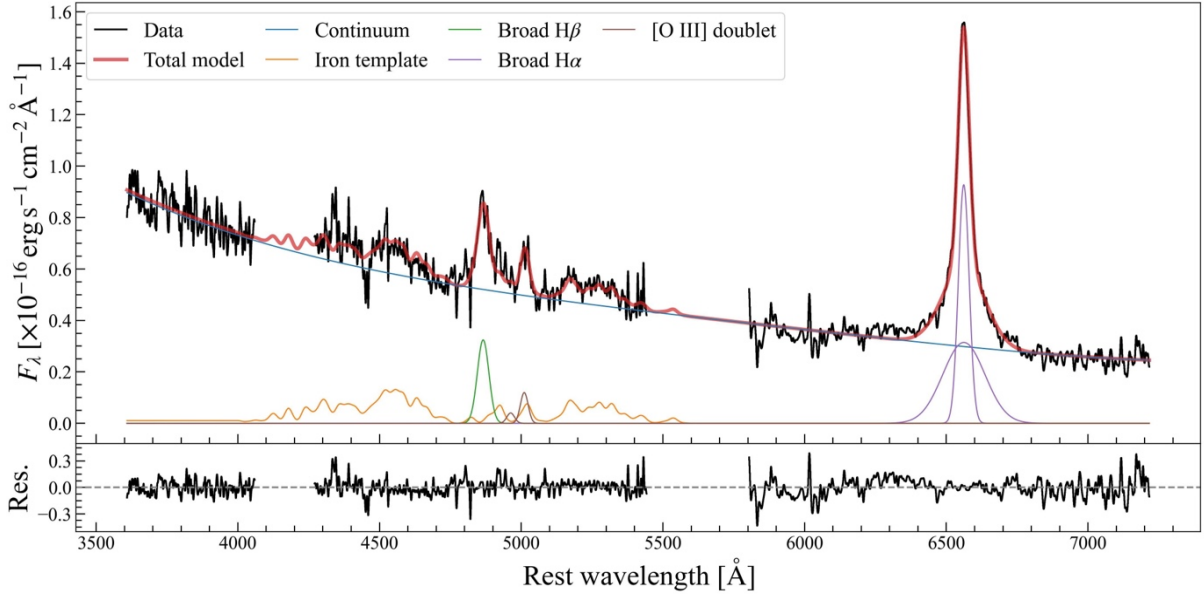

**Extended Data Figure 3 | APO/TripleSpec observed H+K spectrum and spectral decomposition.** The top panel shows our flux calibrated APO/TripleSpec spectrum (black line) together with our best fit model (red line). The model consists of the following components: fourth order polynomial for the continuum (blue line), Fe II template (orange line), Hβ Gaussian emission line (green line), [OIII] Gaussian emission lines (brown lines), and two Hα Gaussian components (purple lines). The best fit matches the data very well with relative residuals (lower panel) below 20%. The data and residual are smoothed by a Gaussian kernel with the standard deviation of three channels for clarity of display, while the fitting is conducted with the originally reduced data.

| PARAMETER/PROPERTY | BEST FIT VALUE |
|---|---|
| **REDSHIFT** | 2.3255 +/- 0.0002 |
| **FWHM(Hα)** | 2526 +/- 65 km s$^{-1}$ |
| **log($L_{Hα}$)** | 45.18 +/- 0.07 erg s$^{-1}$ |
| **FWHM(Hβ)** | 2967 +/- 283 km s$^{-1}$ |
| **EW(Hβ)** | 31.75 +/- 2.56 Angstrom |
| **FWHM([OIII])** | 1926 +/- 266 km s$^{-1}$ |
| **EW([OIII])** | 8.20 +/- 1.29 Angstrom |
| **EW(FEII)** | 38.61 +/- 3.61 Angstrom |
| **R$_{FE}$** | 1.22 +/- 0.11 |
| log($L_{5100}$) | 46.56 +/- 0.07 erg s$^{-1}$ |
| log($L_{BOL}$) | 47.2 – 47.9 erg s$^{-1}$ |

**Extended Data Table 2 | TripleSpec spectral decomposition.**

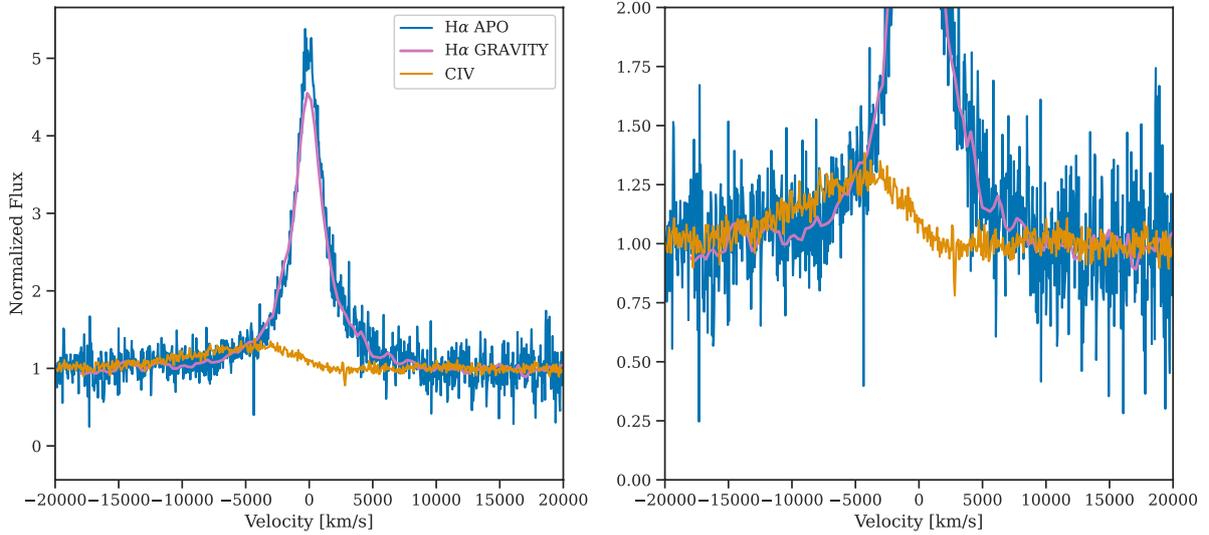

**Extended Data Figure 4. Comparison between CIV and Hα line profiles** We compare the continuum normalized line profiles of CIV from the LAMOST quasar survey (orange) to both of our Hα line profiles from GRAVITY (purple) and APO/TripleSpec (blue). Wavelengths were converted to velocities using the measured redshift of Hα ($z$ = 2.325). CIV shows both a systematic blueshift of ~7000 km s$^{-1}$ and increased line width compared to Hα along with a heavy skew to blueshifted velocities. CIV therefore is likely dominated by outflowing gas and not the virial motion of the BLR.

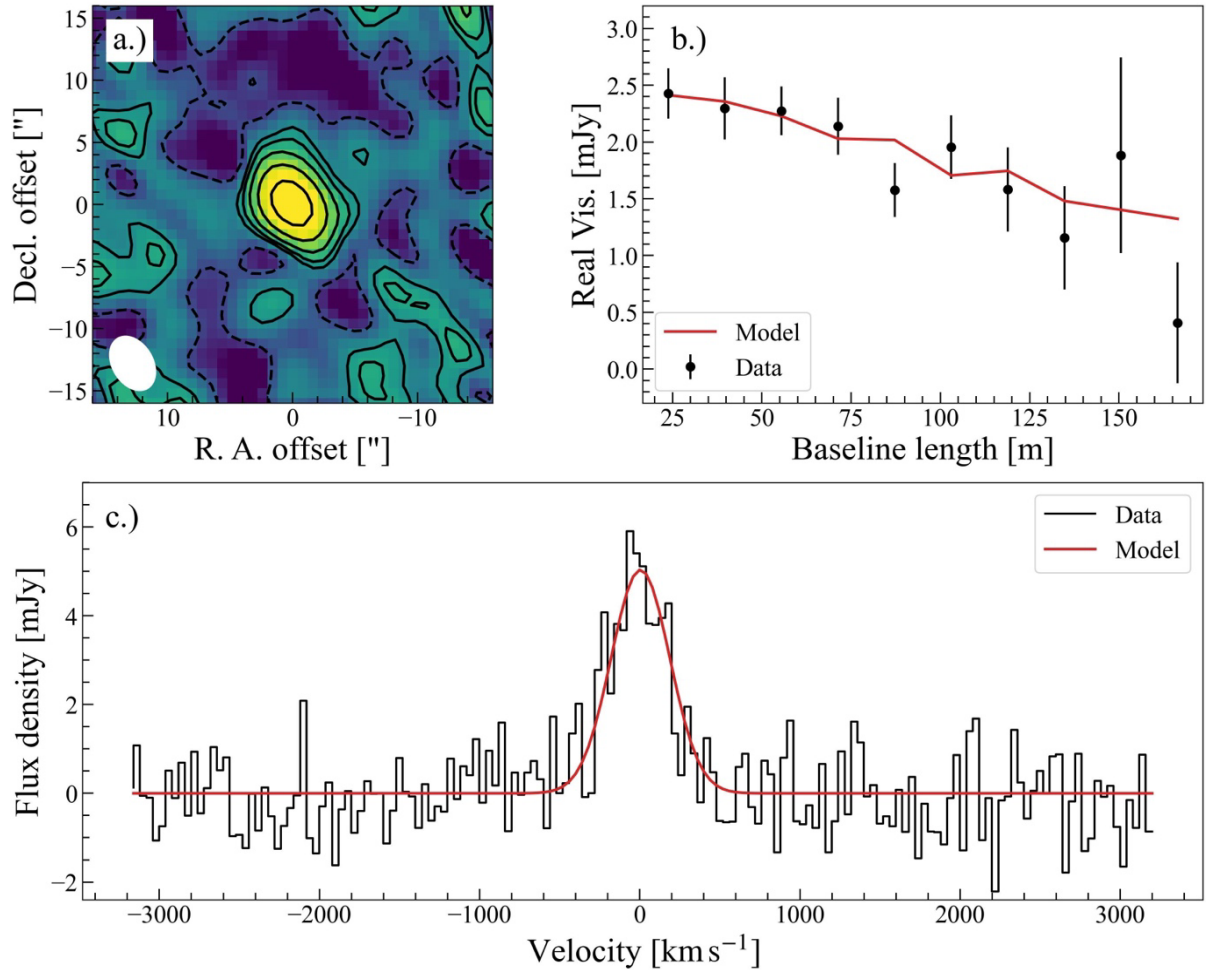

**Extended Data Figure 5. NOEMA CO (3-2) Data and Analysis. a.)** Moment 0 map of J0920 using the channels spanning -700 to 700 km s$^{-1}$ around the expected location of the CO (3-2) line. The contours are (-1, 1, 2, 4, 8, 16) times the RMS noise level, with the -1σ level in the dashed line. The synthesized beam (4.7"x3.2") is shown in the lower left corner. **b.)** Average real part of the visibilities as a function of baseline length (black points) showing decreasing visibility with increasing baseline with 1σ error bars. This indicates J0920 is extended even with the relatively large beam size. The red line is a fit using an elliptical exponential disk model where we find an effective radius of 8.23 kpc. **c.)** Integrated spectrum within the 1σ contour shown in the top left panel showing the detection of the CO (3-2) line. We fit the line with a single Gaussian (red line) finding a FWHM of 432+/- 42 km s$^{-1}$ and use this with the effective radius determined in the top right panel to estimate the dynamical mass of J0920 and place it on the SMBH-galaxy scaling relation (see Main Text).